CORRESPONDENCE

# The Complexity of Synaptic Transmission Revealed by a Multiscale Analysis Approach From The Molecular to The Cellular Level

D. Holcman[1]

Synaptic transmission is significantly reflected in the synaptic current, which depends on several processes such as the location of a released vesicle, the number and type of receptors, trafficking between the postsynaptic density (PSD) and extrasynaptic compartments, as well as the synapse organization. Variations in vesicular release locations, receptor distribution or synaptic geometry modulate the postsynaptic current, making a synapse an intrinsic unreliable device. In their recent Review article, Claire Ribrault et al. propose to couple some of these properties starting from a molecular level to tackle the synaptic variability and they concluded that "the coupling between successive steps in synaptic transmission affects the propagation of fluctuations and therefore may alter the weight of these fluctuations" ("From the stochasticity of molecular processes to the variability of synaptic transmission"Nature Reviews Neuroscience 12, 375-387), suggesting finally that assembling the various steps remains a task to be done.

In this correspondence, we point out recent efforts, using quantitative approaches (biophysical modeling, mathematical analysis and numerical simulations) that were recently used to dissect and integrate different steps of synaptic transmission [1,2,3]. First, it is true that thinking biological processes at molecular level requires a stochastic description [4,5], leading in general to non intuitive behaviors. Integrating the synaptic processes across scales from molecular to cellular is a novel emerging effort of quantitative cellular biophysics. Indeed, through recent modeling, it is now possible to give a biophysical ground for the mean, the variance and the coefficient of Variation (CV), allowing precise computations and studying its dependency as a function of various geometrical parameters such as the PSD size [2]. Interestingly, changing the size of the PSD, while maintaining the number of receptors constant leads to significant changes in the synaptic current, a property already suggested in [6]. Numerical simulations can account for receptor organization in small clusters, which affects the postsynaptic current [12,2,3]. However, the results of all simulation models depend significantly on the receptor properties, usually modeled as Markov chains, capturing the main features of the biophysical properties such as the opening, closing and desensitized states [7,8]. Although several models are available for modeling the dynamics of receptors, it remains a challenging question to account for the native heteromeric structure of the receptor and the ensemble of conductance states. Understanding the synaptic current through a modeling reconstruction would greatly benefit from estimating experimentally the AMPA-channel current, when the number of bound glutamate molecules at the single channel is known.

In addition, direct modeling and mathematical analysis [9] have shown that changing the diffusion coefficient of neurotransmitters such as glutamate molecules in the synaptic cleft cannot affect the number of bound receptors. In less than few microseconds, most of the neurotransmitters have left the synaptic cleft, which is 10 times faster than the receptor dynamical response. Thus any fluctuations leading to a change in the diffusion constant by a factor two or three is unlikely to affect the synaptic current, challenging the conclusion that decreasing the diffusion coefficient could alter the postsynaptic response, as reviewed here in Ribrault et al. (Nature Reviews Neuroscience 12, 375-387). Thus, changing the diffusion coefficient is unlikely to be a source of current fluctuation. The most sensitive parameters in controlling the synaptic current remains the vesicular release location, leading to a continuous range of amplitude current, which can be divided by 8 during an ectopic release [3]. However, when the Active Zone (AZ) and the PSD are aligned, this situation generates a maximal current, when the other parameters are fixed [3]. Altogether this result suggests that any mechanism that would force vesicles to be released in the AZ will lead to a maximal current, although for that position the CV is not minimal [2].

---

[1] Group of Computational Biology and Applied Mathematics, Institute for Biology, IBENS, Ecole Normale Supérieure, 46 rue d'Ulm 75005 Paris, France.

Another fundamental parameter regulating synaptic transmission is the two-dimensional receptor trafficking and the three-dimensional ionic and molecular motion occurring in dendritic spines, the post-synaptic terminal of many neuronal connections. For example, the spine neck is a key regulator, as quantified in [10], while the entire spine geometry regulates various fluxes, as quantified in [11,12]. To conclude, various key steps in synaptic transmission ranging from molecular to cellular aspects have already been integrated, offering a novel simulation packages. The field of synaptic modeling is now mature enough to offer competitive tools to study synapses, even in pathological conditions or following genetic mutations. For example, it has recently been found that Autism Spectrum Disorders is associated with a shank3 mutation [13], a fundamental molecule of the PSD, which can modify receptor trafficking. It will now be conceivable to use directly computational models to test any possible predictions about the synaptic current, by altering directly in the modeling and simulation, a given molecular pathways, while the synaptic current can be used as a readout. Another example is the mysterious ketogenic diet, well known since biblical time, to lead to a significant decrease of epilepsy crisis. Many pathways remains to be studied, but as recently observed, due to a decay in the number of neurotransmitters in vesicles, the synaptic current is decreased [14]. Using a modeling approach, it is possible to predict precisely from how much this number has decreased and to estimate many others associated parameters such as the vesicular fusion location. Future efforts should also include the role of astrocytes in controlling through glio-transmitters release, the pre- and postsynaptic terminals.